\documentclass[journal ]{new-aiaa}
\usepackage[utf8]{inputenc}
\usepackage{textcomp}

\usepackage{graphicx}
\usepackage{amsmath}
\usepackage{longtable,tabularx}

\usepackage[dvipsnames]{xcolor}

\usepackage{amsfonts}
\usepackage{booktabs}
\def\tsc#1{\csdef{#1}{\textsc{\lowercase{#1}}\xspace}}
\tsc{WGM}
\tsc{QE}
\def\myref{\text{\scriptsize ref}}

\newcommand*\dif{\mathop{}\!\mathrm{d}}  %
\newcommand\eq{\textrm{eq}}
\newcommand\neqi{\textrm{neq}}

\title{ Perfectly Matched Layers and Characteristic Boundaries in Lattice Boltzmann: Accuracy vs. Cost }

\author{Friedemann Klass
        \footnote{Ph.D. Candidate, University of Wuppertal, klass@math.uni-wuppertal.de} 
       }
       \affil{University of Wuppertal, 42119 Wuppertal,~Germany}
\author{Alessandro Gabbana
        \footnote{Postdoctoral Researcher, Los Alamos National Laboratory, agabbana@lanl.gov}
        }
        \affil{CCS-2 Computational Physics and Methods, Los Alamos National Laboratory, Los Alamos, 87545 New Mexico, USA}
        \affil{Center for Nonlinear Studies (CNLS), Los Alamos National Laboratory, Los Alamos, 87545 New Mexico, USA}
\author{Andreas Bartel
        \footnote{Lecturer, University of Wuppertal, bartel@math.uni-wuppertal.de}
        }
        \affil{University of Wuppertal, 42119 Wuppertal,~Germany}

\begin{document}

\maketitle

\begin{abstract}
Artificial boundary conditions (BC) play a ubiquitous role in numerical simulations of transport phenomena 
in several diverse fields, such as fluid dynamics, electromagnetism, acoustics, geophysics and many more.
They are essential for accurately capturing the behavior of physical systems whenever the simulation domain
is truncated for computational efficiency purposes.
Ideally, an artificial BC would allow relevant information to enter or leave the computational domain 
without introducing artifacts or unphysical effects. 
Boundary conditions designed to control spurious wave reflections are referred to as 
non-reflective boundary conditions (NRBC).
Another approach is given by the perfectly matched layers (PML), in which
the computational domain is extended with multiple dampening layers, 
where outgoing waves are absorbed exponentially in time.

In this work, we revise the definition of PML in the context of the 
lattice Boltzmann method. 
We evaluate and compare the impact of adopting different types of BC
at the edge of the dampening zone, both in terms of accuracy and computational costs.
We show that for sufficiently large buffer zones, PML allows stable and accurate 
simulations even when using a simple zero-th order extrapolation BC. 
Moreover, employing PML in combination with NRBC potentially offers
significant gains in accuracy at a modest computational overhead,
provided the parameters of the BC are properly tuned to 
match the properties of the underlying fluid flow.

\end{abstract}

\section{Introduction}\label{sec:intro}
The perfectly matched layer (PML) technique is an established tool to treat artificial boundaries that arise when 
truncating the physical domain of a given problem, allowing to limit the propagation of any unphysical activity in the bulk domain.
Originally developed for the absorption of electromagnetic waves~\cite{berenger-jocp-1994}, the PML introduces 
a dampening zone beyond the artificial boundary in which a modified set of governing equations is solved.
This modification ensures that  i) outgoing waves decay exponentially and ii) the dampening zone is \textit{perfectly} 
matched to the bulk in the sense that the interface between bulk and PML region will not cause reflections.
A major advantage of the PML method is that it allows use of
simple boundary conditions (BCs) beyond the dampening zone, as unphysical reflections 
arising at the boundary are absorbed by the dampening zone before they start interacting with the bulk.
PMLs have been successfully applied  to elastodynamics~\cite{collino-geophysics-2001}, the Schrödinger~\cite{xavier-cicp-2008}, 
Helmholtz~\cite{singer-jocp-2004}, Euler~\cite{hu-jcp-1996, hu-ac-1996, hu-jocp-2001} and Navier-Stokes~\cite{hagstrom-aiaa-2005,hagstrom-aiaa-2007,hu-jocp-2008} equations, 
as well as Boltzmann~\cite{hu-AIAA-conf-2010,sutti-jocp-2024} and lattice Boltzmann equations~\cite{tekitek-cmwa-2009,najafi-comflu-2012,shao-jtca-2018}.
In addition to PML, other buffer-zone-like techniques have been developed for flow simulations, 
such as solution filtering in exit zones to reduce boundary reflections~\cite{poinsot-jocp-1992,colonius-aiaa-1993,freund-aiaa-1997}.

In the context of the lattice Boltzmann method (LBM) the definition of BC can represent a complex task, since they
must be specified at the mesoscopic level, in terms of discrete particle distribution functions.
Due to its simple implementation, PML is commonly employed in LBM simulations, in particular in 
the simulation of radiation~\cite{chen-pre-2020} and acoustic phenomena~\cite{rajamuni-preprint-2023,chen-jocp-2023}.

The LBM considered in this work is based on high order velocity stencils which support simulations 
of thermal compressible flow~\cite{shan-jofm-2006,shan-pre-2010}. 
These kind of lattices exhibit multiple layers of nodes, which further complicate the task of 
defining boundary conditions in terms of unknown values of (discrete) particle distribution functions.
Under these settings, the conceptual advantage of the PML to enable use of simple BCs is even more appealing.
Regardless of the order of the velocity stencils used, the dampening zone introduced in the method increases 
the computational cost, since the stream and collide paradigm of the LBM needs to be applied on an extended grid. 
This motivates the question whether the thickness of the dampening zone required to obtain a given level of accuracy 
can be minimized by employing non-reflecting BCs (NRBC) beyond the dampening zone.
While the combination of PML and NRBC has been reported before~\cite{chen-jocp-2023}, a thorough analysis investigating 
the trade-off between accuracy and computational costs is still missing.
In this work we present an implementation of the PML combined with characteristic boundary conditions 
(CBC)~\cite{poinsot-jocp-1992} for the modelling of artificial boundaries. 
We compare this approach to the classical setup present 
in the literature \cite{najafi-comflu-2012}, where the simple Zero Gradient (ZG) BC is used beyond the dampening zone.
For both of these implementations, we evaluate the impact of the width of the dampening zone on both 
the obtained accuracy and the computational cost. 
We show that the combination of characteristic BCs with PML can provide significant gains in terms of accuracy, 
and at relatively modest computational overhead, provided that the parameters of both CBC and PML are carefully tuned. 
Moreover, when the accuracy provided by the ZG BC beyond the PML is sufficient, 
then the ZG BC offers simpler implementation and faster evaluation, even when using high-order multi-speed LBM models.

This work is structured as follows:
Sec.~\ref{sec:lbm} contains a brief description of the LBM. 
In Sec.~\ref{sec:PML}, we revisit in details the implementation of the PML, and its coupling with non-reflective BC. 
Subsequently, we present the BCs, which we combine with PML in order to enhance the boundary treatment in Sec.~\ref{sec:bcs}.
Then, in Sec.~\ref{sec:numerics} we discuss numerical results in which we evaluate the tread-off between accuracy and
computational costs of different types of BC for a few standard numerical benchmarks. 
Finally, in Sec.~\ref{sec:conclusions} we summarize our results and discuss possible future developments.

\section{Lattice Boltzmann Method}\label{sec:lbm}
The LBM~\cite{kruger-book-2017, succi-book-2018} is an established computational fluid dynamic solvers which, 
unlike traditional methods which explicitly discretize the Navier Stokes equations, makes use of a mesoscopic
representation, solving a minimal version of the Boltzmann equation discretized on a Cartesian grid. 

The description of the fluid fields is given in terms of a set of 
$q$ discrete velocity distribution functions $f_i(\mathbf{x}, t)=f(\mathbf{x}, \mathbf{c}_i, t)$ 
(lattice populations hereafter), giving the probability of finding a particle with (discrete) velocity $\mathbf{c}_i$ 
at the node of a Cartesian grid $\mathbf{x}$ at time $t$. 

The time evolution of the populations is governed by the discrete lattice Boltzmann-BGK equation
\begin{equation}\label{eq:lbe}
  f_i(\mathbf{x}+ \mathbf{c}_{i} \Delta t , t + \Delta t) 
  =  
  f_i(\mathbf{x},t) + \Omega_i^{BGK} \quad i = 1,\ \dots,\ q,
\end{equation} 
where collisions are modeled using the single-relaxation time Bhatnagar-Gross-Krook (BGK) model~\cite{bhatnagar-pr-1954}
\begin{equation*}
    \Omega_i^{BGK} = - \frac{1}{\tau} \left( f_i(\mathbf{x},t) - f_{i}^\eq (\mathbf{x},t) \right) ,
\end{equation*}
where populations relax towards their local equilibrium state $f_{i}^\eq$ at rate $\tau$. 

The macroscopic hydrodynamic quantities are obtained from the velocity moments of the distribution.  
Working in rescaled units, where the Boltzmann constant $k_b$ and the molecular mass $m$ are set
to unity, density $\rho$, velocity $\mathbf{u}$ and internal energy density $\rho e$ are given as 
\begin{align}\label{eq:macro}
  \rho            = \sum\limits_{i=1}^q f_i                                             ,  \;\;\quad     
  \rho \mathbf{u} = \sum\limits_{i=1}^q f_i \mathbf{c}_{i}                              ,  \;\;\quad
  d \rho e        = \sum\limits_{i=1}^q f_i \vert \mathbf{c}_{i} - \mathbf{u}\vert^2    ,
\end{align}
where $d$ is the number of spatial dimensions.
We consider here a mono-atomic gas and an ideal equation of state, 
thus the temperature $T$, internal energy $e$ and hydrodynamic pressure $P$ are put in relationship via~\cite{shan-jofm-2006}
\begin{equation}
    e = \frac{d}{2} T , \quad P = \rho T c_s^2 .
\end{equation}

Using a multi-scale Chapman-Enskog expansion~\cite{chapman-book-1990} it is possible to establish a link
between the mesoscopic parameters and the kinematic viscosity $\nu$ %
\begin{equation*} %
  \nu = \left( \tau - \frac{1}{2} \right) c_s^2 ,
\end{equation*}
with $c_s$ the lattice speed of sound.

\begin{figure*}
    \centering
    \includegraphics[width=\linewidth]{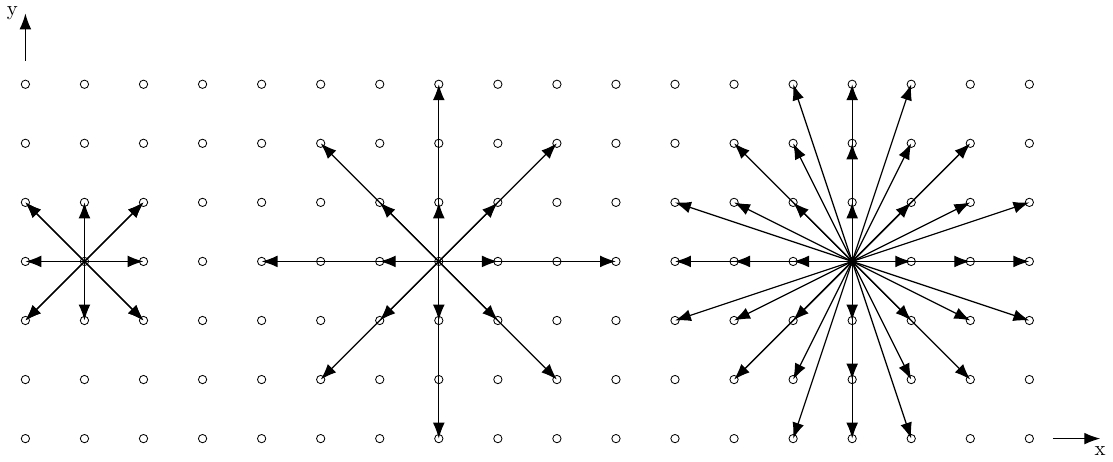}
    \caption{Discrete velocities for the D2Q9 
             (left), D2Q17 (center) and D2Q37 (right) velocity stencils.
            }\label{fig1:stencils}
\end{figure*}

\subsection*{Velocity stencils}
To ensure that Eq.~\eqref{eq:macro} holds, the discrete velocity stencil needs to be chosen according to the abscissa 
of a sufficiently high-order Gauss-Hermite quadrature~\cite{shan-jocs-2016}
i.e., $ \{ ( \omega_i, \mathbf{c}_{i} ) : \, i=1,\, \ldots,\, q \}$, where $\omega_i$ are the quadrature weights. 
Different velocity stencils are commonly referred to using 
the D$d$Q$q$ nomenclature, where $d$ refers to the number of spatial dimensions and $q$ to
the number of discrete velocities.

In what follows we will restrict ourselves to the two-dimensional $(d = 2)$ case, and
consider the D2Q9, D2Q17 and D2Q37 stencils depicted in Fig.~\ref{fig1:stencils}.
While the commonly adopted D2Q9 model captures density and velocity, the
underlying quadrature is not accurate enough to describe the temperature field~\cite{philippi-pre-2006,shan-pre-2010}.
The D2Q17 velocity stencil can correctly recover the third order velocity moments of the particle distribution, 
however, it fails to capture the non-equilibrium component of the heat-flux~\cite{shan-jofm-2006},
making it suitable only for simulations of mildly compressible flow.
In contrast, the D2Q37 stencil enables simulations of fully compressible thermal flows.
For the sake of computational efficiency, it is of interest to minimize the amount of discrete velocities.
The velocity stencils here considered consist of the minimal amount of velocities required to 
form fifth-, seventh- and ninth-order quadrature rules, when additionally, non-negative weights 
and integer-valued velocities with a maximum displacement of $\mathcal{M}=3$ are demanded~\cite{shan-pre-2010}.

An explicit expression for the equilibrium is obtained from an expansion in Hermite polynomials of 
the Maxwell-Boltzmann distribution~\cite{shan-jofm-2006}.
Following the discussion in Ref.~\cite{shan-pre-2010}, we consider a second order expansion for the D2Q9 stencil,
\begin{equation}\label{eq:feq-2}
    f^{\eq,2}_{i}   (\rho, \mathbf{u}, T)   = \,  \omega_i \rho  
    \bigg( 
        1 + \mathbf{u} \cdot \mathbf{c}_{i}  
          + \frac{1}{2c_s^2} 
        \big[
             (\mathbf{u} \cdot \mathbf{c}_{i})^2 - u^2 + (T-1) (c_i^2  - 2)
        \big]
    \bigg),
\end{equation} 
a third order expansion for the D2Q17 stencil,
\begin{align}\label{eq:feq-3}
    f^{\eq,3}_{i}   (\rho, \mathbf{u}, T)   
    = 
    \, f^{\eq,2}_{i}(\rho, \mathbf{u}, T) + \omega_i \rho  \frac{\mathbf{u} \cdot \mathbf{c}_{i}}{6c_s^4}
    \bigg(          
         (\mathbf{u} \cdot \mathbf{c}_{i})^2 - 3  u^2 + 3(T-1) (c_i^2  - 4)
    \bigg),
\end{align} 
and a fourth order expansion for the D2Q37 stencil
\begin{align}\label{eq:feq-4} 
        f^{\rm{eq},4}_{i} & (\rho, \mathbf{u},T)   = \, f^{\eq,3}_{i}(\rho, \mathbf{u}, T)   +   \omega_i \rho   
            \frac{1}{24 c_s^6} 
               \bigg(     (\mathbf{u} \cdot \mathbf{c}_{i})^4 
                      - 6 (\mathbf{u} \cdot \mathbf{c}_{i})^2  u^2 + 3  u^4
                      + 6 (T-1) 
               \big(      (\mathbf{u} \cdot \mathbf{c}_{i})^2 ( c_{i}^2 -  4)
                    + \lvert \mathbf{u} \rvert^2 (4- c_{i}^2) 
               \big)    \notag & \\
        & \hspace{35ex}  
         + 3 (T-1)^2 ( c_{i}^4 - 8  c_{i}^2 +  8)
               \bigg) ,
\end{align} 
respectively, where  $u^2= \mathbf{u} \cdot \mathbf{u}, 
c_i^2= \mathbf{c}_i \cdot \mathbf{c}_i $ and the speed of sound $c_s$ 
is a lattice specific constant.

We conclude this section by outlining the algorithm for solving Eq.~\ref{eq:lbe}.
First, the  the discrete distribution $f_i$ are initialized, for example according to the equilibria
(Eq.~\eqref{eq:feq-2}-\eqref{eq:feq-4}) evaluated at desired macroscopic fields.
Each time step consists of 1) evaluation of the collision operator (right hand side of Eq.~\ref{eq:lbe}) and 
2) streaming step in which populations are moved to neighboring nodes according to the directions defined by the
velocity stencil.
Unless working on a fully periodic domain, the streaming step will leave post-streaming populations at boundary nodes 
unspecified, thus requiring employing suitable BC for defining missing populations before updating the macroscopic 
values by means of Eq.~\eqref{eq:macro} and proceeding to the next iteration. 

\section{Perfectly Matched Layers}\label{sec:PML}

\begin{figure*}[htb]
    \centering
    \includegraphics[width=0.8\linewidth]{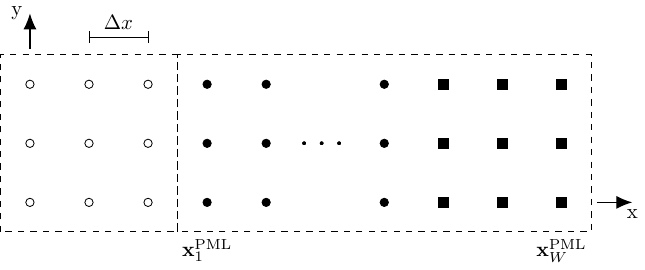}
    \caption{Sketch of PML geometry for a right-hand side boundary on an equidistant grid. 
            }\label{fig:PML}
\end{figure*}

In this section we summarize the derivation of PML for LBM, following the approach originally presented in Ref.~\cite{najafi-comflu-2012}.
The general idea of PML consists in extending the computational domain introducing a dampening layer
where governing equations are modified such that reflection waves are damped out and (ideally) no reflections arise at the interface.
In Fig.~\ref{fig:PML} we sketch the PML geometry for a right-hand side boundary. In the figure, 
the bulk region (hollow nodes) is padded with $W$ layers (filled nodes), where a modified governing equation is solved.
The flow in the absorbing zone is dampened towards a mean equilibrium state, in turn enabling use 
of standard boundary conditions at the edge of the computational domain (square nodes).

For the implementation of this concept at the in the mesoscale LBM framework, the starting point is 
the splitting of the particle distribution $f_i$ as the sum of an equilibrium term $f_{i}^{\eq}$ and a 
non-equilibrium component $f_{i}^{\neqi}$
 \begin{equation}\label{eq:decomp-NJ}
    f_i =  f_i^{\eq} + f_i^{\neqi} =  \bar{f_i}^{\eq} + \hat{f_i}^{\eq} + f_i^{\neqi}, 
\end{equation}
where $f_i^{\eq}$ is further decomposed into a (global) mean equilibrium component $\bar{f_i}^{\eq}$ and a
deviation from the mean equilibrium $\hat{f_i}^{\eq}$. 
The PML approach aims at suppressing the fluctuations contained in $\hat{f_i}^{\eq}$.
That is, the flow is driven towards a prescribed mean equilibrium state $\bar{f}^\eq$ in the dampening region, 
absorbing reflections occurring at boundary nodes. 

Starting from the discrete velocity Boltzmann-BGK equation in two spatial dimensions
\begin{equation}\label{eq:BE-BGK-discrete-vel-PML} 
    \frac{\partial f_i}{\partial t} + c_{i,x}  \frac{\partial f_i}{\partial x} + c_{i,y}  \frac{\partial f_i}{\partial y} 
    =  
    \Omega_i^{BGK},   
\end{equation}
we introduce non-negative dampening coefficients~\cite{collino-geophysics-2001,hu-jocp-2001}, 
$\sigma_x$ and $\sigma_y$ respectively along the $x$ and $y$ component, 
which in combination with Eq.~\ref{eq:decomp-NJ} allow to cast the LHS of Eq.~\ref{eq:BE-BGK-discrete-vel-PML} as
\begin{equation}\label{eq:pml-general}
    \frac{\partial \hat{f}_i^{\eq}}{\partial t} + c_{i,x}  \frac{\partial \hat{f}_i^{\eq}}{\partial x} +
     c_{i,y}  \frac{\partial \hat{f}_i^{\eq}}{\partial y}  
    =
     \underbrace{- \left[  
     \sigma_y c_{i,x} \frac{\partial Q_i} {\partial x}   \
     + \sigma_x c_{i,y} \frac{\partial Q_i} {\partial y}
     + ( \sigma_x + \sigma_y) \hat{f}_i^{\eq} 
     + \sigma_x \sigma_y Q_i 
     \right]}_{:= \Omega_i^{\rm PML}} ,
\end{equation}
with $Q_i$ defined as
\begin{equation} \label{eq:PMLQ}
    \frac{\partial Q_i}{\partial t} = \hat{f}_{i}^{\eq}.
\end{equation}
It follows that due to the decomposition in Eq.~\ref{eq:decomp-NJ} the PML can be expressed as a modification 
of the collision term:
\begin{equation}\label{eq:BE_PML}
    \frac{\partial f_i}{\partial t} + c_{i,x}  \frac{\partial f_i}{\partial x} +c_{i,y}  \frac{\partial f_i}{\partial y} 
    =  
    \Omega_i^{BGK} + \Omega_i^{\rm PML}.
\end{equation}

Hereafter we assume $\sigma_x = \sigma_y = \sigma$. 
This choice, which allows to simplify $\Omega_i^{\rm PML}$ down to
\begin{equation}\label{eq:PMLCOLL}
    \Omega_i^{\rm PML} = - \sigma \left( \mathbf{c}_{i} \cdot \nabla Q_i + 2 \hat{f}_i^{\eq} + \sigma Q_i \right) ,
\end{equation}
proves to be beneficial in terms of numerical stability ~\cite{najafi-comflu-2012}.
The absorption coefficient $\sigma$ controls the strength of the dampening effect of the PML.
We remark that $\sigma$ is not necessarily a constant but may vary spatially and in time.
In this work, we restrict our analysis to one specific case, and follow the common practice~\cite{hu-AIAA-conf-2010} 
of ramping up the absorption coefficient $\sigma$ quadratically with respect to the distance from the interface.
Formally, the dampening strength in the $i$-th level of nodes in the PML (cf. Fig.~\ref{fig:PML}) is given by
\begin{equation}\label{eq:PML_ramp}
    \sigma(\mathbf{x}^{\mathrm{PML}}_i) = \sigma_{\max} \left( \frac{i}{W} \right)^2, \quad i=1, \, 2, \,  \ldots, W,
\end{equation}
where $W$ is the depth of the PML.
Therefore, in our analysis $W$ and $\sigma_{\max}$ will be the free parameters of the PML, which can be tuned
for optimized accuracy and stability~\cite{modave-nme-2014}.

After discretization of space and time, the implementation of the PML for LBM is as follows.
In the bulk, the usual lattice Boltzmann-BGK equation Eq.~\eqref{eq:lbe} is evolved.
In the dampening zone, the collision operator is modified by adding Eq.~\eqref{eq:PMLCOLL}, thus the governing equation 
reads as
\begin{equation}
  f_i(\mathbf{x}+ \mathbf{c}_{i} \Delta t , t + \Delta t) 
  =  
  f_i(\mathbf{x},t) + \Omega_i^{BGK}(\mathbf{x},t)+ \Omega_i^{\rm PML}(\mathbf{x},t).
\end{equation}
We use the trapezoidal rule to evaluate the quantity $Q_i$ in Eq.~\eqref{eq:PMLQ}
\begin{equation*}
    Q_{i,t+\Delta t} 
    = 
    \frac{1}{2} \left( \hat{f}^{\eq}_{i,\Delta t} + f^{\eq}_{i,t+\Delta t} - \bar{f}^{\eq}_{i,t+\Delta t} \right) ,
\end{equation*}
while its spatial gradients are evaluated using second order finite differences, specifically we use
one-sided differences at the outermost layer of nodes $\mathbf{x}^{\mathrm{PML}}_W$
\begin{align}\label{eq:FD-dx-bw}
    \frac{\partial Q_i(\mathbf{x}_{b,M})} {\partial \alpha} 
    \approx
    \tfrac{ 1}{2} 
    \bigl(  
            3 Q_i( \mathbf{x}) 
            \!-\! 4 Q_i(\mathbf{x} - \Delta x \ \mathbf{e}_\alpha) 
               +  Q_i(\mathbf{x} - 2 \Delta x \ \mathbf{e}_\alpha)    
    \bigr) .
\end{align}
while central finite differences are used for all other cases
\begin{align}\label{eq:FD-dx-center}
    \frac{\partial Q_i(\mathbf{x}_{b,j})} {\partial \alpha}  
    \approx 
    \tfrac{ 1}{2} 
    \bigl(  
               Q_i(\mathbf{x} +  \Delta x \ \mathbf{e}_\alpha)
            -  Q_i(\mathbf{x} -  \Delta x \ \mathbf{e}_\alpha)
    \bigr),  
\end{align} 
where $\alpha$ is the spatial index and $\mathbf{e}_\alpha$ the outward facing normal vector corresponding to the boundary.

We stress once again that the action of the PML in LBM is entirely contained in the modified collision operator appearing 
in Eq.~\eqref{eq:PMLCOLL}, thus preserving the the stream and collide paradigm of the numerical method.

\section{Boundary Conditions}\label{sec:bcs}
The implementation of the PML described in the previous section still requires that a boundary condition is specified
at the outer-most layer of the absorbing zone.
Since the PML absorbs reflections occurring at the boundaries, in principle even very crude and simple BCs 
can be employed in combination with PML.
However, the grid extension introduced by the dampening zone leads to increase in computational costs, and we are 
here interested in investigating whether the choice of the BC applied at the edge of the dampening zone can 
be used to minimize the PML width $W$, still yielding a desired level of accuracy.
In our analysis we will consider two types of BC, respectively a simple zero-gradient BC and a non-reflective BC,
which are below detailed.

\subsection*{Zero Gradient}
The zero-gradient BC (ZG) represents arguably the simplest possible choice for an outflow BC.
Here populations at the boundary nodes $\mathbf{x}_{b,j}, \  j= 1, \,2, \,\dotsc, \mathcal{M}$ are obtained with
a zero-th order extrapolation from the adjacent fluid node $\mathbf{x}_f$:
\begin{equation}\label{eq:zg-bc}
    f_i(\mathbf{x}_{b,j},t+\Delta t) = f_i(\mathbf{x}_f,t+\Delta t)   \quad i =1,\; 2, \; \ldots,\; q.
\end{equation}

\subsection*{Characteristic BC}
We consider recently developed characteristic BC~\cite{klass-cicp-2023,klass-camwa-2024} (CBC) for LBM, 
which allows to modulate incoming wave amplitudes by considering an hyperbolic approximation of the target governing macroscopic equations.
We summarize below the two main ingredients in the derivation of CBC.

\subsubsection*{Characteristic Analysis}
Starting from a macroscopic transport equation (Navier-Stokes-Fourier in our case, providing the time evolution
of the macroscopic fields $\mathbb{U}:=\left( \rho, u_x, u_y, T \right)^\top$), one aims at casting the equation 
in the following form
\begin{equation}\label{eq:NSF-CHAR}
  \frac{\partial \mathbb{U}}{\partial t} 
  = 
  - A \frac{\partial \mathbb{U}}{\partial x} + \mathbb{T} + \mathbb{V}.
\end{equation}
In the above, $\mathbb{T}$ and $\mathbb{V}$ represent transversal and viscous contributions.
Upon discarding these contributions, Eq.\ref{eq:NSF-CHAR} simplifies down to the so-called  
locally one dimensional inviscid (LODI) approximation, which allows for analyzing the 
amplitude variations of waves crossing the boundary. 
This analysis can be performed by first diagonalizing $A = S^{-1} \Lambda S$, and next by
identifying the amplitude variations of waves crossing the boundary via
\begin{equation}
  \mathcal{L}_x
  = 
  \begin{pmatrix}
    \mathcal{L}_{x,1}, \ \mathcal{L}_{x,2} , \ \mathcal{L}_{x,3}, \ \mathcal{L}_{x,4}
  \end{pmatrix}^\top 
  = 
  \Lambda S \frac{\partial \mathbb{U}}{\partial x},
\end{equation} 
where the sign of $\Lambda_{ii}$ determines the propagation direction the $i$-th wave. 
Thus, it is possible to distinguish between waves entering and leaving the computational domain.
This makes it possible to replace the vector $\mathcal{L}_x$ with a modified vector $\mathcal{\bar{L}}_x$, 
where amplitudes of outgoing waves are retained and incoming wave amplitudes are manipulated.
\subsubsection*{Variations of characteristic BCs}
Several variations of characteristic BCs can be identified depending on how incoming wave amplitudes 
$\mathcal{\bar{L}}_{x,\mathrm{inc}}$ are treated.
In a perfectly non-reflecting boundary condition, they are simply set to zero~\cite{thompson-jocp-1987},
\begin{equation}\label{eq:PNR}
    \mathcal{\bar{L}}_{x,\mathrm{inc}} = 0.
\end{equation}
However, this in practice often leads to a drift in the macroscopic target quantities in the long term.
Imposing a relaxation of target macroscopic quantities towards a reference state~\cite{yoo-ctan-2005}
offers a (partial) remedy to the problem.

Once $\mathcal{\bar{L}}_x$ is specified, the hyperbolic approximation to Eq.~\eqref{eq:NSF-CHAR} reads~\cite{jung-jocp-2015}
\begin{equation}\label{eq:thermal_CBC_macro_evo}
    \frac{ \partial \mathbb{U}}{ \partial t} 
    = 
    -S^{-1}  \mathcal{\bar{L}}_x + \mathbb{T} + \mathbb{V} 
    = 
    -S^{-1}  (\mathcal{\bar{L}}_x + \mathcal{T} + \mathcal{V} )   ,
\end{equation}

In the spirit of the LODI approximation used to compute $\mathcal{\bar{L}}$, the so-called LODI BC 
is obtained by dropping transversal and viscous contributions~\cite{heubes-jcam-2014}, where the macroscopic system reads
\begin{equation}\label{eq:thermal_LODI_macro_evo}
    \frac{ \partial \mathbb{U}}{ \partial t} 
    = 
    -S^{-1}  \mathcal{\bar{L}}_x.  
\end{equation}

Numerically solving either Eq.~\eqref{eq:thermal_CBC_macro_evo} or \eqref{eq:thermal_LODI_macro_evo} 
yields macroscopic target values $\mathbb{U}_\mathrm{tgt}$ (see Ref.~\cite{klass-camwa-2024} for details).

Finally, it is possible to use $\mathbb{U}_{tgt}$ to implement a mesoscopic Dirichlet BC.
We consider here the non-equilibrium extrapolation method 
\begin{equation*}
    f_i(\mathbf{x}_{b,j}, t +\Delta t ) 
    = 
    f_{i}^{\eq} (\rho_{\mathrm{tgt}}, \mathbf{u}_{\mathrm{tgt}}, T_{\mathrm{tgt}}) + f_{i}^{\neqi} (\mathbf{x}_f,  t +\Delta t)
\end{equation*}
where the non-equilibrium part of populations at the boundary nodes $\mathbf{x}_{b,j}, \  j= 1, \,2, \,\dotsc, \mathcal{M}$ 
is copied from the nearest fluid node $\mathbf{x}_f$ (see e.g. Ref.~\cite{klass-jocs-2021} for a discussion on
other possible choices for multi-speed LBM).

Finally, we remark that it is in principle possible to include the extra terms introduced by the PML in the 
characteristic analysis (see Appendix~\ref{sec:appendix-pml-cbc} for details).

\section{Numerical Results}\label{sec:numerics}
In this section we report our numerical results, evaluating accuracy and computational costs of employing
PML in combination with different underlying BC, also varying the free parameters offered by the PML/BC.
We consider two standard numerical benchmarks often employed in the evaluation of artificial 
boundaries~\cite{klass-camwa-2024,jung-jocp-2015,chen-jocp-2023}, namely a mono-dimensional type of flow 
triggered by an initial density step, in Sec.~\ref{subsec:1dstep}, and the propagation of a vortex out of 
the computational domain in Sec.~\ref{subsec:vortex}.
We start by detailing the performance metrics and how we established reference data.

\subsection*{Reference Simulation}
The accuracy of the BCs is quantified by computing errors with respect to reference fields. 
Such reference fields $Z^{\text{ref}}$ are obtained from a fully periodic LBM simulation 
on an extended grid for the quantities $Z \in \{ \rho, u_x, u_y, T \}$. 
The reference simulation is conducted on a grid sufficiently large such that no interaction takes 
place between the boundaries and the bulk dynamics in the region of interest, ensuring that  
the only possible source of error lies in the boundary treatment.

In what follows we will report global relative $L^2$-errors $e_Z$ and pointwise relative $L^2$-errors $\widetilde{e}_Z$ 
with respect to the reference fields, which we compute as
\begin{align*}
    e_Z = \left( \sum_{(x,y)\in L_x\times L_y} 
                  \left( \frac{| Z(x,y) - Z^{\myref}(x,y)|}{|Z^{\myref}(x,y)|} 
                  \right)^{\!\! 2} 
          \right)^{\frac{1}{2}}\!, 
    \quad
    \widetilde{e}_Z(x,y) = \frac{|Z(x,y) - Z^\myref(x,y)|}{ |Z^\myref(x,y)|}.
\end{align*}
Moreover, we also define an average global error $\bar{e}_Z$, computed by averaging $e_Z$ over all the iterations sampled. 
Finally, we also introduce the quantity $C_{Z}$ to compare the average error of a specific implementation against
the one obtained employing the ZG BC:
\begin{equation}\label{eq:RuntimeVsAccMetric2}
  C_{Z}= \frac{ \bar{e}_{Z}}{\bar{ e}_{Z}^{\rm ZG}}, \quad Z \in \{ \rho, u_x, u_y, T \} .
\end{equation}

\subsection*{Computational Cost}
In a typical LBM simulation, the cost of evaluation the BC at the edges of the computational domains is typically 
negligible with respect to the total execution time, specially for large grid sizes.
In order to establish a fair comparison between the different BC schemes, in our analysis we measure the (wall clock) 
time $t^{\rm BC}$ spent in computing kernels devoted to the processing of boundary conditions. 
If the PML is used, this will also includes the time required for the additional stream 
collide operations in the dampening zone.
In what follows we will indicate with $C_t$ the relative increase in the cost of processing boundary nodes with
respect to execution time required for evaluating the ZG BC, i.e.
\begin{equation}\label{eq:RuntimeVsAccMetric1}   
  C_t = \frac{t^{\rm BC}}{t^{\rm ZG}}. 
\end{equation}

\subsection{One-dimensional density step test case}\label{subsec:1dstep}
We consider a flow with a one-dimensional dynamic, with initial conditions given by a
homogeneous background velocity $\mathbf{u}_0 = (u_x, 0)$ (parallel to the $x$-axis) and a smooth density step given by
\begin{align}
    \rho(x,y) = \! \begin{cases}
        \rho_1 + \frac{\rho_1-\rho_0}{2}  \left( \tanh \left( \text{s} (x-\tfrac{1}{4} L_x )  \right) -1\right), \qquad \text{ if } x \leq \frac{L_x}{2},  \\
        \rho_1 - \frac{\rho_1-\rho_0}{2}  \left( \tanh \left( \text{s} (x-\tfrac{3}{4} L_x )  \right) -1\right), \qquad \text{ else},
    \end{cases} %
\end{align}
where the parameter $s$ controls the steepness of the initial step.
We perform simulations on a grid of size $L_x \times L_y = 200\times 20$. %
Numerical values for the parameters are $\rho_0=1, \rho_1=1.05, s=0.5$ and the background velocity 
is determined from the Mach number, which is here set to $\mathrm{Ma}=0.05$.

The left- and right-hand side boundaries of the rectangular domain are equipped with artificial BC, 
while upper and lower boundaries are periodic. 
In this benchmark, the mean equilibrium $\bar{f}^\eq$ used in the PML domain is computed with respect 
to the quantities $\rho_0, \mathbf{u}_0$. The perfectly non-reflecting LODI BC based on Eqs.~\eqref{eq:PNR} 
and ~\eqref{eq:thermal_LODI_macro_evo} is used as a characteristic BC.
\begin{figure}[h]
  \centering
  \includegraphics[width=\linewidth]{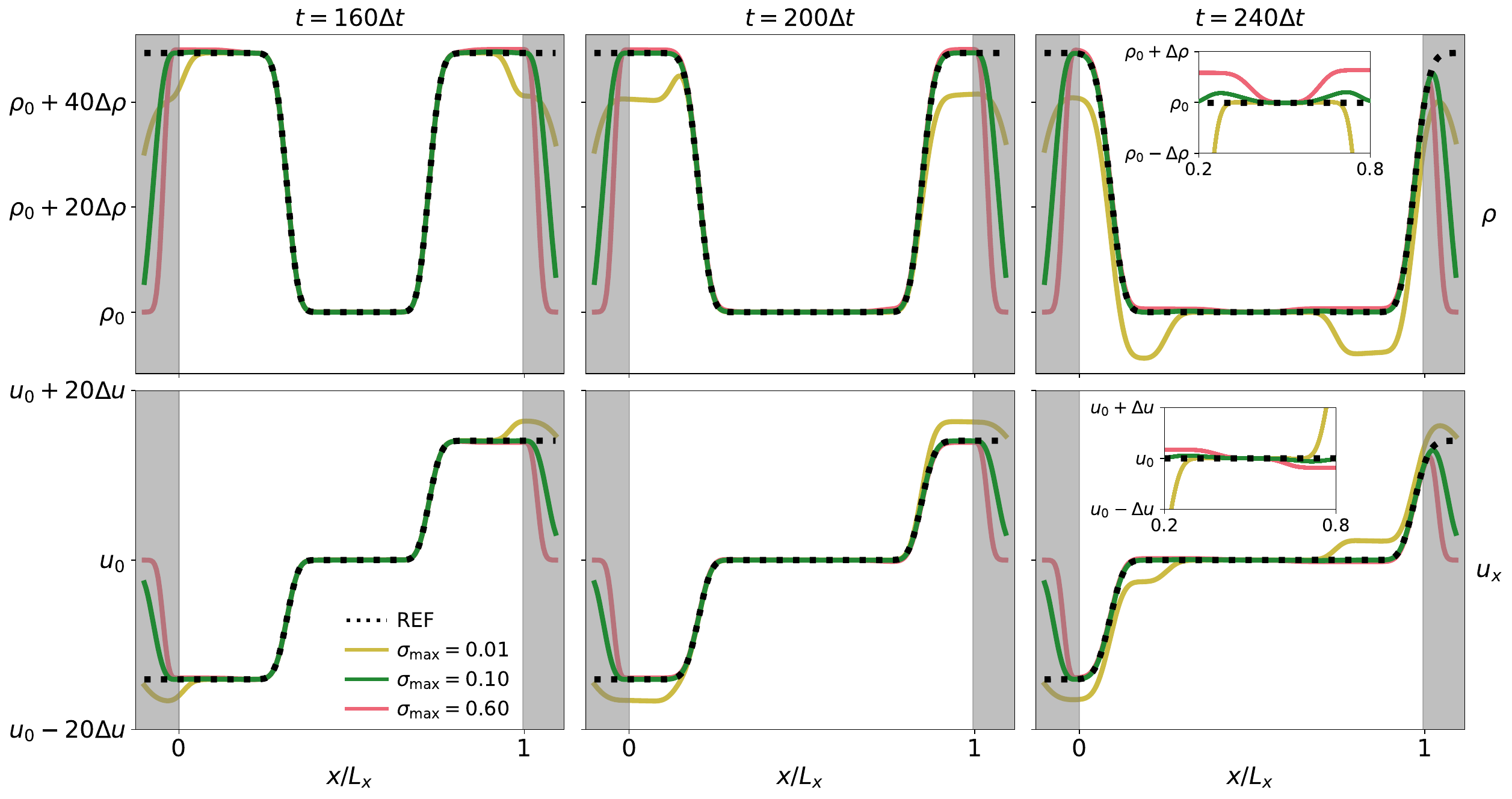}
  \caption{ Time evolution of the density $\rho$ and $x$-component of the velocity field $u_x$ for the ``density step'' 
            test case. 
            Numerical values for the axis extent are $\Delta \rho = 0.0005, \Delta u = 0.001$. 
          }\label{fig:isothermal-step-profiles}
\end{figure}

The effect of the dampening zone on the flow fields can be observed from Fig.~\ref{fig:isothermal-step-profiles}, where
profiles of the density and stream-wise velocity are shown for selected time steps along the horizontal midplane 
$y = L_y / 2$, and the ZG BC is used beyond the PML.
In the figure, the gray-colored areas at the left and right hand of the domain represent $W=20$ additional 
layers of nodes used by the PML. We compare the reference solution (dotted lines), obtained on an extended grid,
with simulations on the reduced domain using different values of $\sigma_{\max} = \{ 0.01, 0.10, 0.60 \}$
to highlight the importance of carefully selecting the value of the dampening rate. 
When $\sigma_{\max}$ is too small ($\sigma_{\max}=0.01$ in our example), the dampening effect is not sufficiently 
strong to absorb the reflections originating from the ZG BC, which are seen polluting the bulk domain.
On the other hand, taking a larger value $\sigma_{\max}=0.60$ leads to a more rapid decay towards the macroscopic 
values imposed by $\bar{f}^\eq$ in the dampening zone. However, this creates a mismatch at the interface with
an overshooting of the macroscopic fields in the bulk of the domain (cf. inset in Fig~\ref{fig:isothermal-step-profiles}).
Finally, we show that it is possible to find an optimal value, $\sigma_{\max}=0.10$, which allows to minimize the
impact of the reflection waves of the bulk dynamic.

\begin{figure}[h!]
    \centering
    \includegraphics[width=\linewidth]{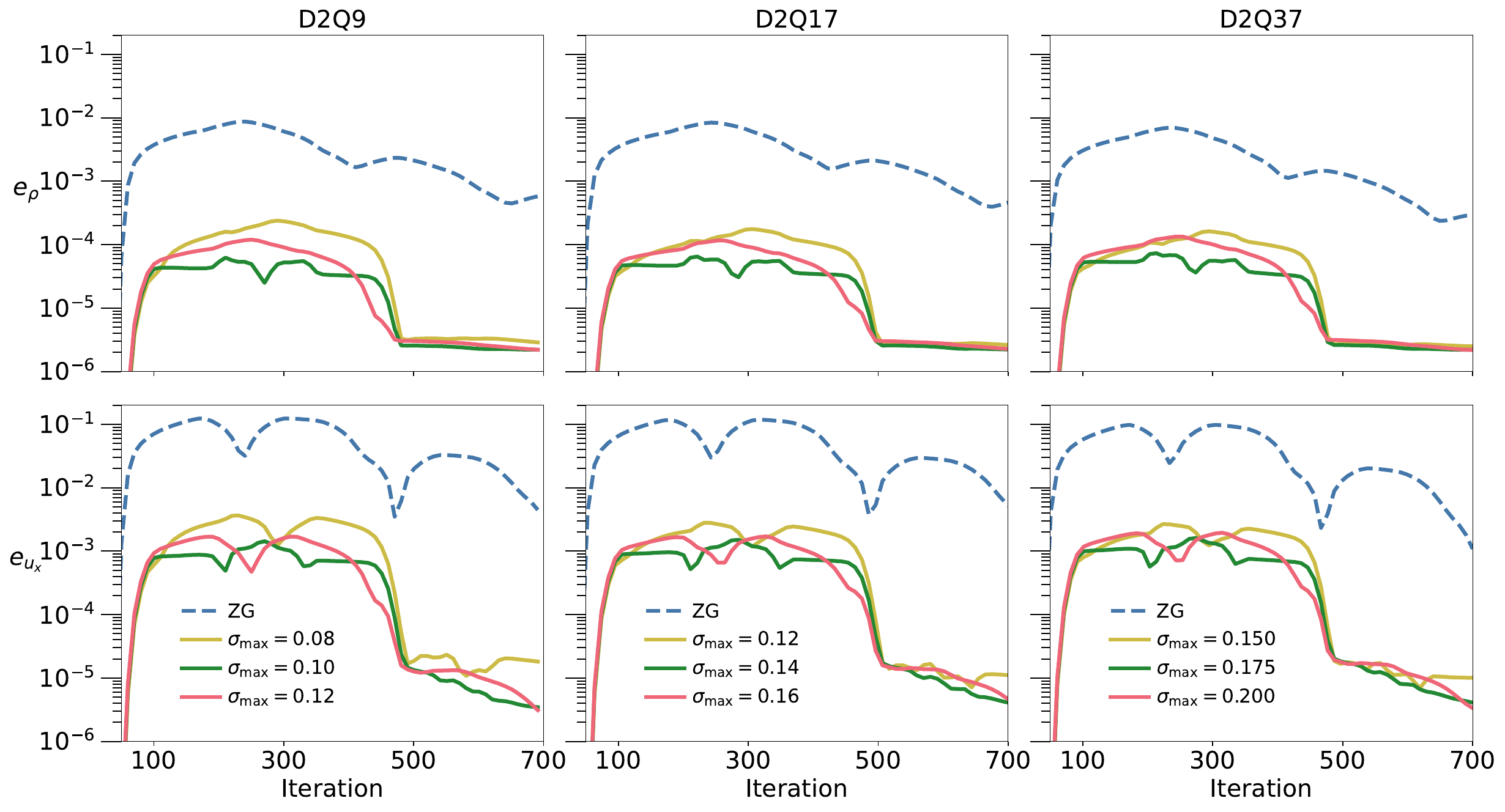}
    \caption{ Time evolution of global errors for density (top row) and velocity (bottom row) for the ``density step''
              benchmark, using ZG BC.
            }\label{fig:scan_sig_W20}
\end{figure}

On a more quantitative ground, in Fig.~\ref{fig:scan_sig_W20} we report the time evolution of 
global relative $L^2$-errors $e_Z$, for the three different stencils employed in this work, and for different 
values of $\sigma_{\max}$.
We show that combining ZG with PML reduces the global errors $e_Z$ by about two orders of magnitude 
over the pure ZG BC without any dampening layers (dashed line in the figure).
The level of accuracy is comparable for both the single speed D2Q9 model, and the multi-speed D2Q17 and D2Q37.
The optimal value of $\sigma_{\max}$, which is critical for overall accuracy, was determined by a parameter scan, and
it is found to depend on the value $W$, the velocity stencil, and on the BC used at the outer-most layer of the PML.
\begin{figure}[h!]
    \centering
    \includegraphics[width=\linewidth]{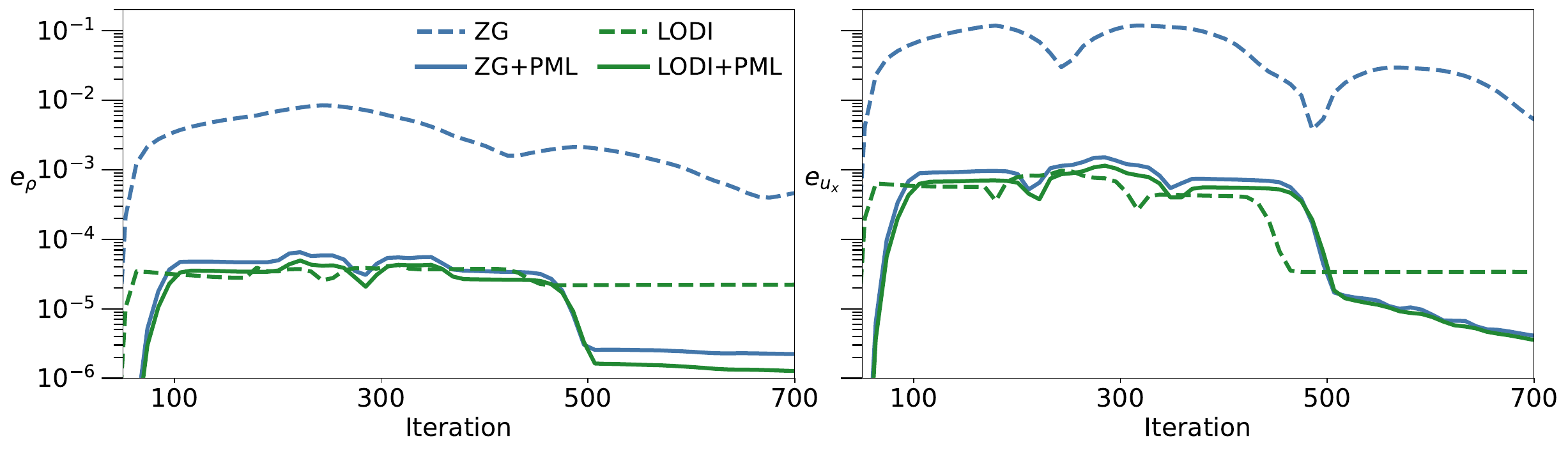}
    \caption{ Time evolution of the global errors for density (left) and velocity (right) fields for the ``density step''
              benchmark, comparing ZG and LODI for the D2Q17 model.
            }\label{fig:L2-PMLCBC}
\end{figure}
In Fig.~\ref{fig:L2-PMLCBC} we report a similar analysis, this time comparing ZG and LODI BC, with (solid lines) and without (dashed lines) PML,
for the D2Q17 velocity stencil. The dampening coefficients used for the data reported in the figure
are respectively $\sigma_{\max}=0.08$ for LODI and $\sigma_{\max}=0.14$ for ZG.

We observe that for a fixed value of $W=20$, the accuracy obtained using the LODI BC 
beyond the PML is approximately the same as for the ZG BC case.
Moreover, the LODI BC without PML is able to match the accuracy of PML (and even minimize the maximum error) 
up until the end of the interaction of the density step with the boundary.
Afterwards, the PML drives the flow to the correct steady state defined by $\bar{f}^\eq(\rho_0, \mathbf{u}_0)$,
yielding improved accuracy over non-PML BCs in the later stages of the numerical simulation.
\begin{figure}[h!]
  \centering
  \includegraphics[width=0.99\linewidth]{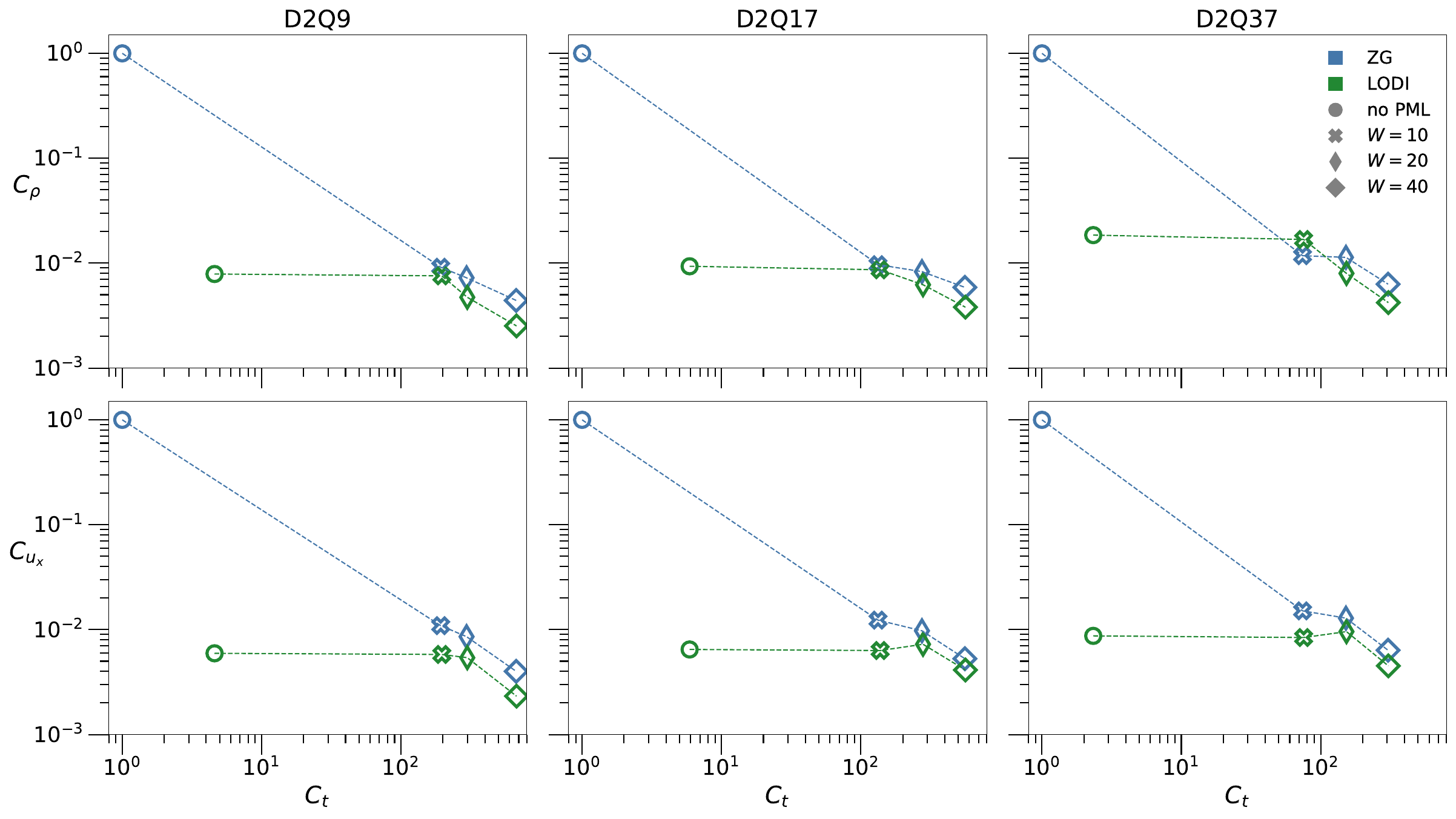}
  \caption{ Computational cost vs accuracy for the ``density step'' test case.
            See text for definition of the error metric $C_Z$ and scaled computational cost $C_t$.
          }\label{fig:isothermal-step-runtime}
\end{figure}

We conclude the analysis for this first benchmark considering the trade-off between computational cost and accuracy 
of the boundary treatment. In Fig.~\ref{fig:isothermal-step-runtime} we report the quantity 
$C_t$ (cf. Eq~\eqref{eq:RuntimeVsAccMetric2}) vs $C_Z$ (cf. Eq.~\eqref{eq:RuntimeVsAccMetric1}),
i.e. the relative global error and relative computational cost of the BC computing kernels scaled with respect to
data for the ZG BC without PML. The values of dampening coefficient $\sigma_{\max}$ used in the different cases are tabulated in Table~\ref{tab:STEP-SIGMAX}.
\begin{table}[h!]
    \caption{Values of $\sigma_{\max}$ used for the ``density step'' test case.
             We report PML-BC combinations with varying PML thickness $W$ in dependence of the underlying 
             discrete velocity stencil. 
            }\label{tab:STEP-SIGMAX}
    \centering
    \begin{tabular}{c ccc ccc ccc }
        \toprule
             & \multicolumn{3}{c}{D2Q9} & \multicolumn{3}{c}{D2Q17} & \multicolumn{3}{c}{D2Q37} \\ 
        \cmidrule(r){1-1} \cmidrule(r){2-4}  \cmidrule(r){5-7} \cmidrule(r){8-10} 
        BC   &  W=10   & W=20  & W=40    &  W=10     & W=20   &  W=40 &  W=10    & W=20     & W=40 \\ 
        \cmidrule(r){1-1} \cmidrule(r){2-4}  \cmidrule(r){5-7} \cmidrule(r){8-10} 
        ZG   &  0.17   & 0.10  & 0.07    &  0.33     & 0.14   &  0.09 &  0.35    & 0.175    & 0.10 \\
        LODI &  0.001  & 0.04  & 0.04    &  0.001    & 0.08   &  0.04 &  0.001   & 0.10     & 0.06 \\
        \bottomrule               
    \end{tabular}
\end{table}

Since the flow in this benchmark presents a strictly one-dimensional dynamic, the assumptions made in the definition 
of the LODI BC are satisfied and allow to deliver a good level of accuracy even without PML
(two orders of magnitude over the ZG BC regardless of the chosen velocity stencil), at a modest computational overhead.
The introduction of a dampening zone of $W=10$ nodes beyond each horizontal boundary in the PML approach 
increases the computational cost of evaluating the boundary nodes by about two order of magnitude, independently
of the underlying velocity stencil. The overhead comes from the need of applying the stream and collide 
paradigm to the PML nodes.
Evolving the fluid in the dampening zone becomes the dominant cost factor of the boundary treatment even 
at moderate thickness values $W$, as for a given value of $W$, the runtime of all PML/BC combinations is comparable.

Employing a thin PML consisting of $W=10$ nodes on each lateral boundary, the ZG BC gives very similar accuracy to the LODI scheme.
Using the LODI scheme in conjunction with PML or a wider PML zone with the ZG BC gives only small improvements 
in accuracy while increasing the simulation time. Thus, for such a simple flow configuration, 
the usage of characteristic BCs without PML appears to be the most efficient choice.

The values of $\sigma_{\max}$ used in the simulations performed to carried the analysis are 
reported in Tab.~\ref{tab:STEP-SIGMAX}, and they show that since the LODI scheme is capable of significantly
damping reflection waves it requires smaller values of $\sigma_{\max}$ than the ZG BC.

\subsection{Propagating vortex test case}\label{subsec:vortex}
We consider a second benchmark, this time with a fully two-dimensional dynamic, to further put under stress the CBC scheme.
We consider the propagation of a thermal vortex in a rectangular computational domain of size $L_x \times L_y$. 
The initialization of the problem is given in terms of normalized spatial coordinates $(x^{\ast},y^{\ast}) \in [-1,1]^2$:
\begin{equation*}
   x^{\ast} = \frac{2(x-1)}{L_x - 1  } -1, \quad y^{\ast} = \frac{2(y-1)}{L_y - 1} -1.
\end{equation*}
The homogeneous background flow fields are given as 
\begin{equation*}
    \rho(x, y) = \rho_0, \quad \mathbf{u}(x, y) = \mathbf{u_0} 
    = 
    \begin{pmatrix}
        u_{0,x} \\ 
              0
    \end{pmatrix}, 
    \quad 
    T(x,y) = T_0
\end{equation*}
and $u_{0,x}$ is determined from the Mach number which is set to $\mathrm{Ma}=0.1$.
\begin{figure}[htp]
  \centering
  \includegraphics[width=0.99\linewidth]{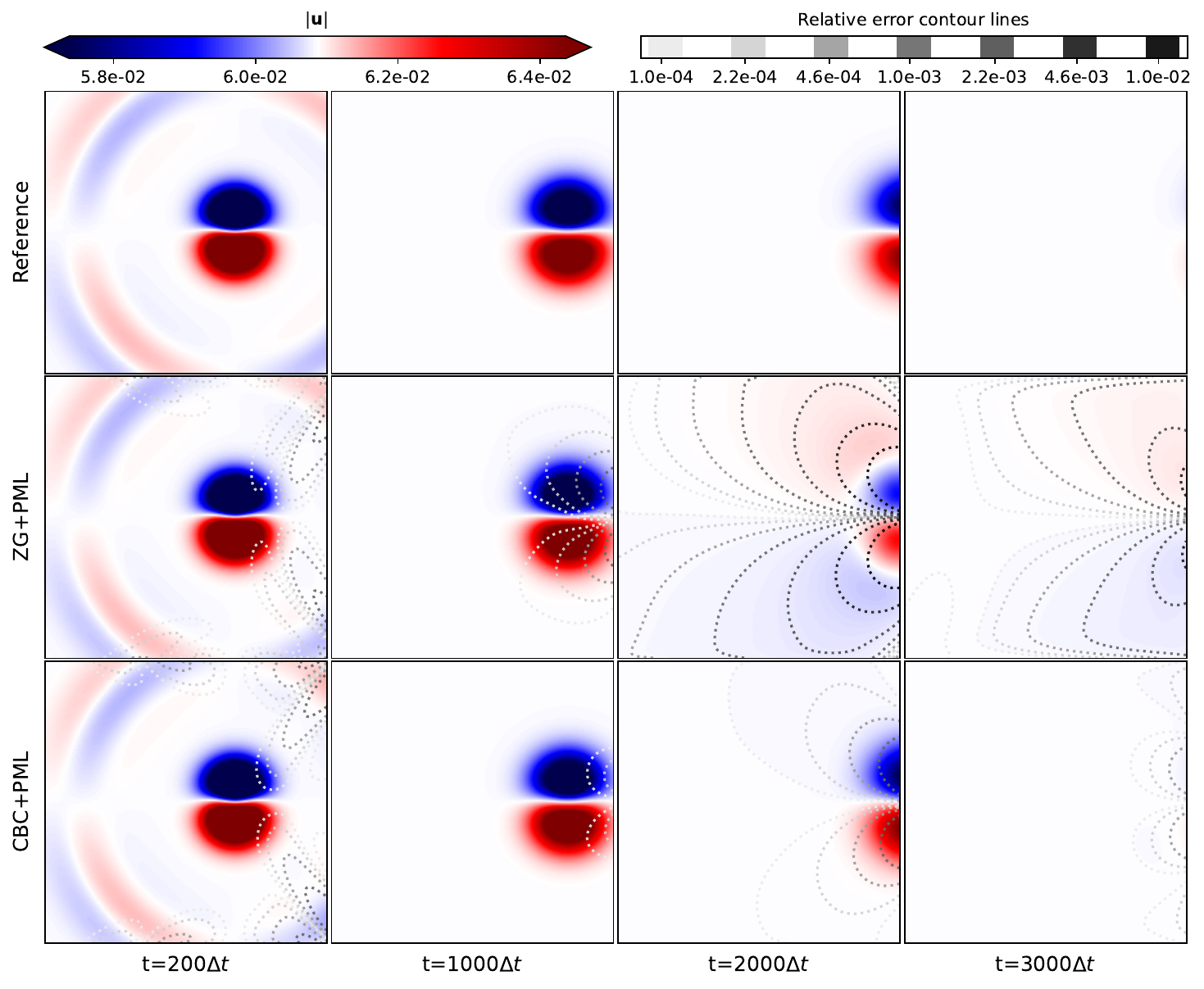}
  \caption{ Time evolution of the magnitude of the velocity field for the thermal vortex benchmark.
            }\label{fig:thermal-vortex-snap-rh}
\end{figure}

The thermal vortex is formed by a perturbation of both temperature and velocity within a circle centered 
at $(x^{\ast}_0, \ 0)$ and with radius $r^{\ast}$. 
The center of the vortex $x^{\ast}_0= \tfrac{K}{L_x - 1}$ is given in terms of a parameter $K$, that defines 
the displacement along the horizontal midplane. More precisely, the initial velocity and temperature fields read 
\begin{equation*}
    \mathbf{u}(x,y) = \mathbf{u_0} + 
    \begin{cases}
        0, & \text{if} \   (x^{\ast}- {x^{\ast}_0)}^2 + {y^{\ast}}^2 \geq {r^{\ast}}^2 
        \\[1ex]
        \mathbf{v}(x^{\ast}- x^{\ast}_0,y^{\ast})   & \text{otherwise},
        \end{cases} \qquad
        T(x,y) = T_0 + 
        \begin{cases}
        0 & \text{if} \   (x^{\ast}- {x^{\ast}_0)}^2 + {y^{\ast}}^2 \geq {r^{\ast}}^2, \\
        \theta(x^{\ast}- x^{\ast}_0,y^{\ast})   & \text{otherwise},
    \end{cases}
\end{equation*}
In the above, the vortex strength in terms of the initial perturbations is given in terms of a shape parameter $b$ as
\begin{align}\label{eq:initialvalues-vortex-example}
     \mathbf{v}(x,y) 
        = 
        \frac{5}{2} u_{0,x} 2^{ -  \frac{x^2 + y^2}{b^2}}
        \begin{pmatrix}    y  \\[1ex] 
        -  x 
        \end{pmatrix}\!, \quad
        \theta(x,y) 
        = 
        y \ \frac{5}{2} u_{0,x} 2^{ -  \frac{x^2 + y^2}{b^2}}. 
\end{align}
The Eckert number corresponding to the simulation parameters is $\mathrm{Ec} \approx 0.02$.

This benchmark is more challenging with respect to the BC, as a flow with fully two-dimensional dynamics is simulated. 
Thus, reflections may occur at an arbitrary angle. Here, all the boundaries are considered as open boundaries and thus, 
the entire computational domain is padded with an absorbing zone of width $W$, giving an effective grid size 
of $(L_x+2W)\times(L_y+2W)$. In what follows we use $L_x = L_y = 300$ and the D2Q17 stencil.

In order to employ characteristic BCs, incoming wave amplitude variations need to be posed at the boundaries. 
Following~\cite{yoo-ctam-2007}, the outlet is treated using the CBC given by Eq.~\eqref{eq:thermal_CBC_macro_evo}, 
where the incoming wave amplitude at the velocity outlet at the right-hand side boundary is posed as
\begin{equation}\label{eq:CBC-Relax}
    \mathcal{\bar{L}}_{x,i} 
    = 
    (1 - \mathrm{Ma} )\mathcal{T}_{x,i}  + \mathcal{V}_{x,i}.                           
\end{equation} 
Other straight edge boundary are treated using the perfectly non-reflecting LODI BC given 
by Eqs.~\eqref{eq:PNR} and \eqref{eq:thermal_LODI_macro_evo}.
Figure~\ref{fig:thermal-vortex-snap-rh} provides snapshots at different time steps of the magnitude of the 
velocity field, comparing the dynamic obtained on a reference grid with that provided by PML with $W=10$ nodes 
in combination with ZG BC (middle row), and with the CBC (bottom row).
The maximum dampening strength was set to $\sigma_{\max}=0.02$ for the ZG BC and $\sigma_{\max}=0.007$ for the CBC.

We observe that right after initialization, a spherical pressure pulse starts traveling through the bulk domain, 
causing interactions with the open boundaries, which are visible already at $t = 200 \Delta t$ (in particular in 
proximity of corner nodes).
After $1000 \Delta t$ iterations, the core of the vortex begins interacting with the boundary. 
From here on, the ZG BC introduces significant reflection waves which pollute the results in the bulk domain.
On the other hand, the entire vortex-boundary interaction is more accurately captured 
if the CBC is used over the ZG BC at the outlet.

\begin{figure}[h!]
  \centering
  \includegraphics[width=0.95\linewidth]{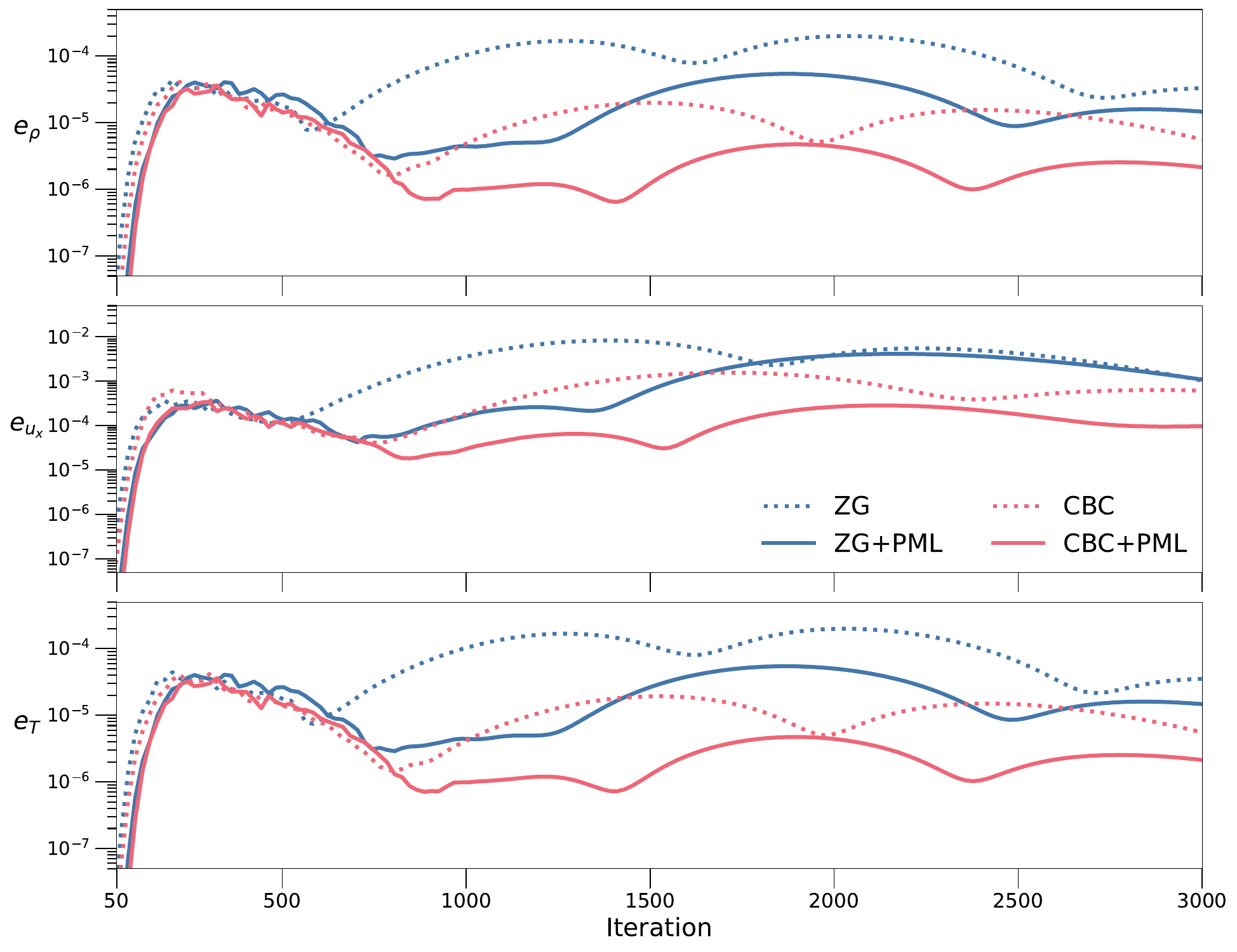}
  \caption{ 
            Time evolution of the global errors for density (top row) $x$-component of the velocity (middle row) 
            and temperature field (bottom row) for the ``thermal vortex'' benchmark.
          }\label{fig:thermal-vortex-L2}
\end{figure}
In Fig.~\ref{fig:thermal-vortex-L2}, the errors caused by the boundary treatment are quantified over the course of the 
simulation by tracking the time evolution of $e_Z$ for different macroscopic fields.
Once again, in the figure we use dashed lines to denote the ZG BC (blue) and the CBC BC (red) without PML,
while continues lines show results for ZG and CBC combined with PML.

As expected, the large errors are observed when using the plain ZG BC without PML.
Combining the ZG BC with a PML of width $W=10$ leads to a level of accuracy comparable 
to that of the characteristic BC without PML.
The accuracy can be further improved by combining CBC with PML.

\begin{figure}[h!]
  \centering
  \includegraphics[width=0.99\linewidth]{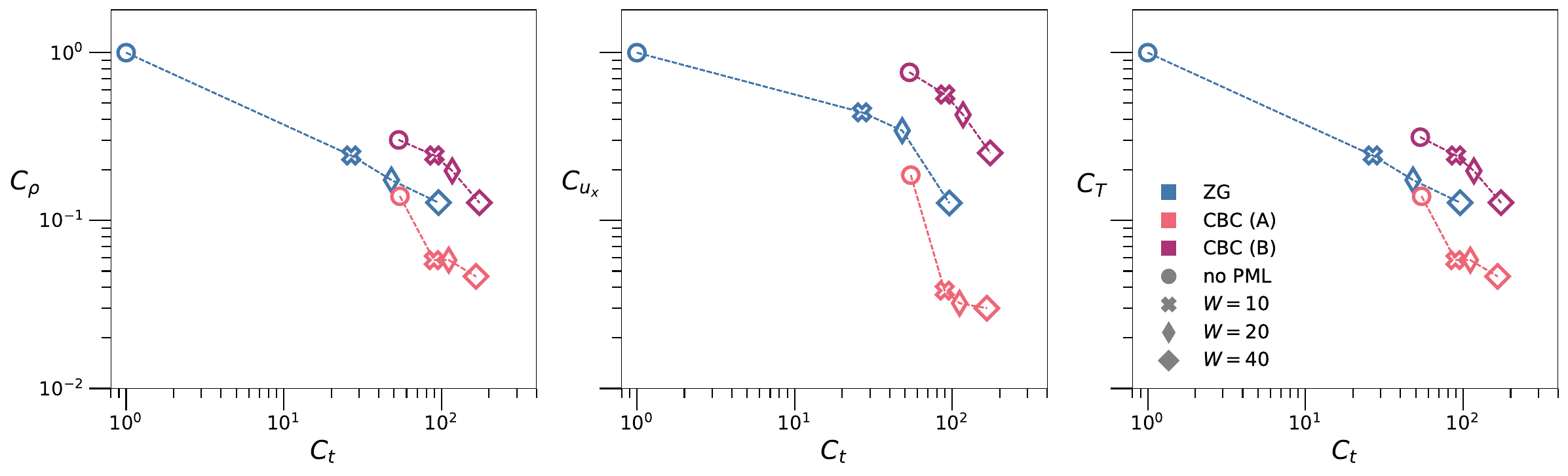}
  \caption{ Computational cost vs accuracy for the ``thermal vortex'' test case.
          }\label{fig:thermal-vortex-runtime}
\end{figure}

The computational cost for this gain in accuracy is inspected in Fig.~\ref{fig:thermal-vortex-runtime}, 
where we present the relative increase in runtime vs the obtained accuracy for the PML used at several values 
of $W$ in conjunction with the various BCs. 
Likewise for the previous benchmark, the optimal value for $\sigma_{\max}$ was obtained by performing a
parameter scan (and the cost of this scan is not taken into account in the analysis).
They are given by, respectively for the ZG (CBC) BC $0.02 (0.007)$ for $W=10$, $0.01 (0.002)$ for $W=20$, and $0.003 (0.001)$ for $W=40$.

In order to evaluate the impact of the choice of how the incoming wave amplitudes are modulated, we compare 
two different CBC schemes. The first one, which has been used so far, and which we here label CBC (A), 
employs the relaxation scheme described by Eq.~\ref{eq:CBC-Relax}. 
The second, CBC (B) uses instead the perfectly non-reflecting version given by Eq.~\ref{eq:PNR}.

For this test-case, the computational time required for applying characteristic BCs increases the time spent 
in the boundary routine by a factor of about 50. This is because all boundaries are taken to be open, 
and the characteristic BCs must also account for the energy equation in this thermal flow configuration, 
which increases the method's overhead compared to simple extrapolation.

As a result, the computational cost for evolving the fluid in the PML region is not as dominant 
as was in the previous benchmark. Using CBC beyond a PML with a width $W=20$ is about as costly 
as using a Zero Gradient (ZG) BC with $W=40$. While CBCs are more expensive than ZG BCs, 
they are still more accurate at a comparable computational cost. 

However, the choice of the parameters of the CBC is critical to preserve the advantages over the ZG BC.
This is shown by comparing the choice for how the incoming wave amplitudes are handled by the two different
CBC schemes. The perfectly non-reflecting CBC (scheme B), where incoming wave amplitudes are set to zero, 
produces less accurate results and requires more runtime than the ZG BC with PML. 
The relaxation of incoming wave amplitudes as described in Eq.~\eqref{eq:CBC-Relax} (scheme A), on the other hand, 
allows to minimize the error for all the macroscopic fields.

In conclusion, the combination of characteristic BCs and PML can be a more efficient choice than the ZG BC 
with PML if a high level of accuracy is required. However, this requires careful tuning of parameters 
for both the PML and the characteristic BC. If the accuracy levels obtained using the ZG BC beyond the PML are sufficient, 
the ZG BC offers a simpler implementation and faster evaluation.

Finally, we note that the outlet boundary is usually the most critical for reducing reflections at open boundaries.
Therefore, the damping zones at the remaining boundaries can be thinner, 
potentially reducing the computational overhead introduced by the PML.

\section{Conclusion}\label{sec:conclusions}
In this paper, we have revised the derivation of perfectly matched layers in the lattice Boltzmann method,
and presented their implementation coupled with a non-reflective boundary condition, supporting both standard
single-speed as well as multi-speed velocity stencils.
We have reported results of a few selected benchmarks to evaluate and compare the 
impact of using different BC in combination with the PML, both in terms of obtained accuracy and computational cost.

We have shown that for simple flow configurations with one-dimensional dynamics, characteristic BCs 
provide the best tread-off between accuracy and computational costs. Since their derivation assumes a locally one-dimensional
dynamic, they allow to dampen the reflection waves introduced by the truncation of the computational domain,
such that their combination with PML does not offer significant gains in terms of accuracy.

On the other hand, when considering flow with truly two-dimensional dynamics, even carefully tuning the parameters
of the CBC leads unavoidably to reflection waves polluting the bulk domain, in turn making the coupling with
PML more compelling.
In this scenario, CBC coupled with PML still provides significant gains in accuracy, and at relatively modest 
computational overhead, than the coupling between a zero-gradient BC with PML, provided that the parameters
of both CBC and PML are carefully tuned. In particular, the performance of the CBC 
critically depends on modeling choice performed for handling the inward pointing wave amplitude variations.

If the accuracy levels obtained using the ZG BC beyond the PML are sufficient, 
the ZG BC offers a simpler implementation, reducing complexity and computational costs even when using higher
order LBM models based on multi-speed stencils.

Finally, the choice of the dampening coefficient $\sigma$ plays a crucial role in the overall accuracy of the PML approach, 
however it is in general not know how this parameter should be chosen in simulations.
Alongside $\sigma$, also the choice of $\bar{f^{\rm eq}}$ might be non trivial, specially in the presence of turbulent flows. 
This, combined with the extra parameters that can be tweaked in the CBC formulation, makes up for a vast search space which can be 
explored in attempt to further improve the performances of this class of NRBC. In a future work we plan to employ a 
data driven approach~\cite{bedrunka-hpc-2021, corbetta-epje-2023} for the optimization of the free parameters of the BC,
and evaluate its effect in the stability of simulations of flows at increasingly large Mach numbers.

\section*{Acknowledgments}
A.G. gratefully acknowledge the support of the U.S. Department of Energy through the LANL/LDRD Program under project 
number 20240740PRD1 and the Center for Non-Linear Studies for this work.

\section*{Appendix}
\renewcommand\thefigure{A.\arabic{figure}} %
\renewcommand{\theequation}{A.\arabic{equation}} %
\setcounter{equation}{0}
\setcounter{figure}{0}

\subsection{Characteristic analysis with the Perfectly Matched Layer}\label{sec:appendix-pml-cbc}

In this Appendix section, we describe the possibility of defining a characteristic BC consistent with the
modified governing equations that arise in the PML approach.

As discussed in the main text, the PML approach for the Lattice Boltzmann Method consists in a modified collision operator
\begin{equation}\label{eq:COLL_APPENDIX}
    \Omega_i = \Omega_i^{BGK} + \Omega_i^{\rm PML},
\end{equation}
where $\Omega_i^{BGK}$ is the BGK-collision operator and the additional term is given by
\begin{equation}\label{eq:PMLCOLL_APPENDIX}
    \Omega_i^{\rm PML} = - \sigma \left( \mathbf{c}_{i} \cdot \nabla Q_i + 2 \hat{f}_i^{\eq} + \sigma Q_i \right).
\end{equation}

In the above, the quantity $\hat{f_i}^{\eq}=f_i^{\eq} - \bar{f_i}^{\eq} $ denotes the deviation from a 
mean equilibrium state $\bar{f_i}^{\eq} = \bar{f_i}^{\eq}(\bar{\rho}, \bar{\mathbf{u}},\bar{T}) $ 
and the term $Q_i$ is calculated from the relation
\begin{equation} \label{eq:PMLQ_APPENDIX}
    \frac{\partial Q_i}{\partial t} = \hat{f}_{i}^{\eq}.
\end{equation}
Thus, usage of PML alters the macroscopic limit of the Lattice Boltzmann Equation by adding 
velocity moments of Eq.~\eqref{eq:PMLCOLL_APPENDIX} to the continuity, momentum and energy equations.
This effect can be expressed in terms of the conserved macroscopic quantities 
$\mathbf{U} = \left( \rho,   \rho \mathbf{u}, \rho e \right)^\top$ and the auxiliary variables 
$\tilde{\mathbf{Q}},\tilde{\mathbf{Q}}_\alpha$, which are given by~\cite{najafi-comflu-2012}
\begin{align*}
    \frac{\partial\tilde{\mathbf{Q}}}{\partial t} 
    = 
    \mathbf{U} - \bar{\mathbf{U}}, \quad \quad \frac{\partial\tilde{\mathbf{Q}}_\alpha}{\partial t} = \int c_\alpha 
    \begin{bmatrix}
        1 \\ c_\beta \\ c_\beta c_\beta
    \end{bmatrix} [f^\eq - \bar{f}^\eq] \dif c,
\end{align*}
where $\bar{\mathbf{U}} = \left( \bar{\rho}, \bar{\rho} \bar{\mathbf{u}}, \bar{\rho} \bar{e} \right)^\top$ 
labels the conserved quantities corresponding to the mean equilibrium $\bar{f}^\eq$.

In conclusion, instead of using Eq.~\eqref{eq:NSF-CHAR} as starting point for the definition of the 
characteristic analysis we can use instead
\begin{equation}\label{eq:NSF-CHAR-PML}
  \frac{\partial \mathbb{U}}{\partial t} 
  = 
  - A \frac{\partial \mathbb{U}}{\partial x} + \mathbb{T} + \mathbb{V} + \mathbb{S} \left(
   \underbrace{\sigma \left( \frac{\partial \tilde{\mathbf{Q}}_x}{\partial x} + \frac{\partial \tilde{\mathbf{Q}}_y}{\partial y} \right)}_{I}
    + \underbrace{2 \sigma \left( \mathbf{U} - \bar{\mathbf{U}} \right) \vphantom{  \resizebox{0.045\hsize}{!}{$\frac{1}{1}$}  }}_{II} + 
    \underbrace{\sigma^2 \tilde{\mathbf{Q}} \vphantom{  \resizebox{0.045\hsize}{!}{$\frac{1}{1}$}  }}_{III} \right),
\end{equation}
where $\mathbb{S} = \frac{1}{\rho} \left( \rho, 1 , \frac{d}{2} \right)^\top$ is a scaling vector used 
to relate primitive to conserved variables.
\begin{figure}[h!]
    \centering
    \includegraphics[width=\linewidth]{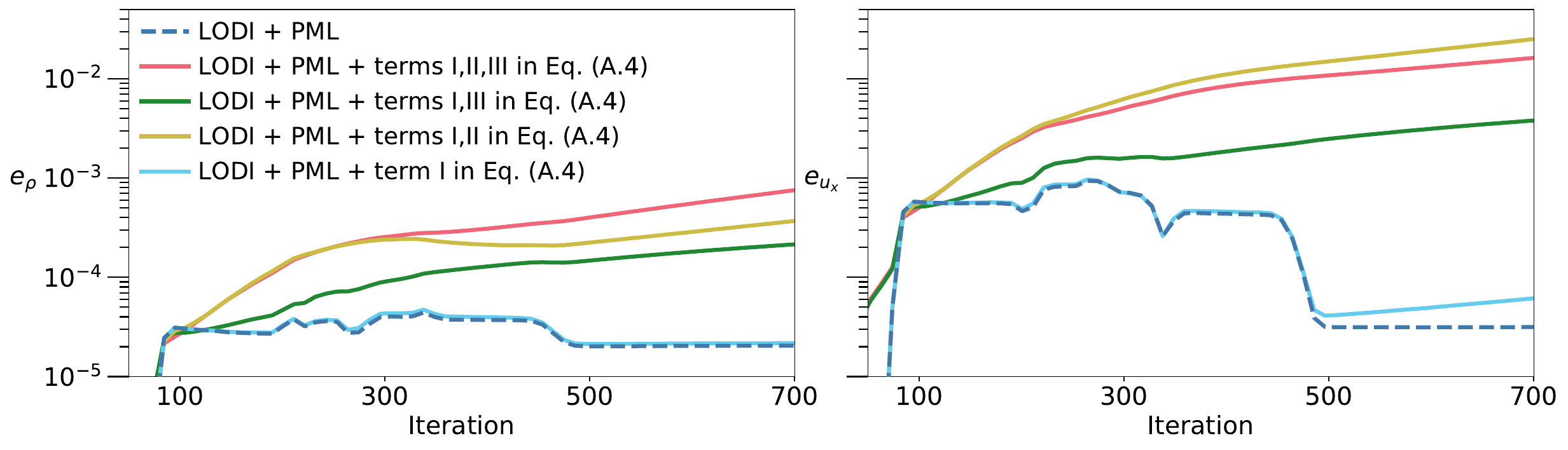}
    \caption{ %
              Evolution of global errors in the isothermal step benchmark on incorporation of the moments 
              of the modified collision operator \eqref{eq:NSF-CHAR-PML} in the LODI BC \eqref{eq:thermal_LODI_macro_evo}. 
            }\label{fig:appendix}
\end{figure}

The characteristic analysis now follows the same steps as described in Sec.~\ref{sec:bcs}, 
the only difference is the presence of the additional source terms.

We consider once again the one-dimensional density step case described in the main text (cf. Sec.~\ref{subsec:1dstep}).
In Fig.~\ref{fig:appendix}, we compare the time evolution of the global error for this new combined LODI-PML,
evaluating the effect of including different combinations of the extra terms present in Eq.\ref{eq:NSF-CHAR-PML}.
All cases make use of a PML of width $W=10$ and $\sigma_{\max}=0.001$.

We observe that this new approach gives comparable accuracy to the combined LODI-PML approach previously presented in
the main text only when discarding the second and third additional terms in Eq.\ref{eq:NSF-CHAR-PML}. 
Other combinations yields a significant decrease in the obtained accuracy.
We speculate that the reason for this comes from the choice of the mean equilibrium state $\bar{f_i}^{\eq}$,
which is here non time/space dependent, and we leave as a open problem the definition of 
an optimal choice for $\bar{f_i}^{\eq}$.

% \bibliography{biblio}

\begin{thebibliography}{33}
\newcommand{\enquote}[1]{``#1''}
\providecommand{\natexlab}[1]{#1}
\providecommand{\url}[1]{\texttt{#1}}
\providecommand{\urlprefix}{URL }
\expandafter\ifx\csname urlstyle\endcsname\relax
  \providecommand{\doi}[1]{\discretionary{}{}{}https://doi.org/#1}\else
  \providecommand{\doi}[1]{\discretionary{}{}{}\urlstyle{rm}\url{https://doi.org/#1}}\fi

\bibitem[{Berenger(1994)}]{berenger-jocp-1994}
Berenger, J.-P., \enquote{{A perfectly matched layer for the absorption of
  electromagnetic waves},} \emph{Journal of Computational Physics}, Vol. 114,
  No.~2, 1994, pp. 185--200.
\newblock \doi{10.1006/jcph.1994.1159}.

\bibitem[{Collino and Tsogka(2001)}]{collino-geophysics-2001}
Collino, F., and Tsogka, C., \enquote{{Application of the perfectly matched
  absorbing layer model to the linear elastodynamic problem in anisotropic
  heterogeneous media},} \emph{GEOPHYSICS}, Vol.~66, No.~1, 2001, pp. 294--307.
\newblock \doi{10.1190/1.1444908}.

\bibitem[{Antoine et~al.(2008)Antoine, Arnold, Besse, Ehrhardt, and
  Sch{\"a}dle}]{xavier-cicp-2008}
Antoine, X., Arnold, A., Besse, C., Ehrhardt, M., and Sch{\"a}dle, A.,
  \enquote{{A Review of Transparent and Artificial Boundary Conditions
  Techniques for Linear and Nonlinear Schr\"{o}dinger Equations},}
  \emph{Communications in Computational Physics}, Vol.~4, 2008, pp. 729--796.
\newblock
  \urlprefix\url{http://global-sci.org/intro/article_detail/cicp/7814.html}.

\bibitem[{Singer and Turkel(2004)}]{singer-jocp-2004}
Singer, I., and Turkel, E., \enquote{{A perfectly matched layer for the
  Helmholtz equation in a semi-infinite strip},} \emph{Journal of Computational
  Physics}, Vol. 201, No.~2, 2004, pp. 439--465.
\newblock \doi{10.1016/j.jcp.2004.06.010}.

\bibitem[{Hu(1996{\natexlab{a}})}]{hu-jcp-1996}
Hu, F.~Q., \enquote{On Absorbing Boundary Conditions for Linearized Euler
  Equations by a Perfectly Matched Layer,} \emph{Journal of Computational
  Physics}, Vol. 129, No.~1, 1996{\natexlab{a}}, pp. 201--219.
\newblock \doi{10.1006/jcph.1996.0244}.

\bibitem[{Hu(1996{\natexlab{b}})}]{hu-ac-1996}
Hu, F., \enquote{On perfectly matched layer as an absorbing boundary
  condition,} \emph{Aeroacoustics Conference}, 1996{\natexlab{b}}, p. 1664.
\newblock \doi{10.2514/6.1996-1664}.

\bibitem[{Hu(2001)}]{hu-jocp-2001}
Hu, F.~Q., \enquote{{A Stable, Perfectly Matched Layer for Linearized Euler
  Equations in Unsplit Physical Variables},} \emph{Journal of Computational
  Physics}, Vol. 173, No.~2, 2001, pp. 455--480.
\newblock \doi{10.1006/jcph.2001.6887}.

\bibitem[{Hagstrom et~al.(2005)Hagstrom, Goodrich, Nazarov, and
  Dodson}]{hagstrom-aiaa-2005}
Hagstrom, T., Goodrich, J., Nazarov, I., and Dodson, C., \enquote{High-order
  methods and boundary conditions for simulating subsonic flows,} \emph{11th
  AIAA/CEAS Aeroacoustics Conference}, 2005, p. 2869.
\newblock \doi{10.2514/6.2005-2869}.

\bibitem[{Hagstrom and Appelo(2007)}]{hagstrom-aiaa-2007}
Hagstrom, T., and Appelo, D., \enquote{Experiments with Hermite methods for
  simulating compressible flows: Runge-Kutta time-stepping and absorbing
  layers,} \emph{13th AIAA/CEAS Aeroacoustics Conference (28th AIAA
  Aeroacoustics Conference)}, 2007, p. 3505.
\newblock \doi{10.2514/6.2007-3505}.

\bibitem[{Hu et~al.(2008)Hu, Li, and Lin}]{hu-jocp-2008}
Hu, F.~Q., Li, X., and Lin, D., \enquote{{Absorbing boundary conditions for
  nonlinear Euler and Navier\textendash}{S}tokes equations based on the
  perfectly matched layer technique,} \emph{Journal of Computational Physics},
  Vol. 227, No.~9, 2008, pp. 4398--4424.
\newblock \doi{10.1016/j.jcp.2008.01.010}.

\bibitem[{Hu and Craig(2010)}]{hu-AIAA-conf-2010}
Hu, F.~Q., and Craig, E., \enquote{{On the Perfectly Matched Layer for the
  Boltzmann-BGK Equation and its Application to Computational Aeroacoustics},}
  \emph{16th {AIAA}/{CEAS} Aeroacoustics Conference}, American Institute of
  Aeronautics and Astronautics, 2010.
\newblock \doi{10.2514/6.2010-3935}.

\bibitem[{Sutti and Hesthaven(2024)}]{sutti-jocp-2024}
Sutti, M., and Hesthaven, J.~S., \enquote{Perfectly matched layers for the
  Boltzmann equation: Stability and sensitivity analysis,} \emph{Journal of
  Computational Physics}, Vol. 509, 2024, p. 113047.
\newblock \doi{10.1016/j.jcp.2024.113047}.

\bibitem[{Tekitek et~al.(2009)Tekitek, Bouzidi, Dubois, and
  Lallemand}]{tekitek-cmwa-2009}
Tekitek, M., Bouzidi, M., Dubois, F., and Lallemand, P., \enquote{{Towards
  perfectly matching layers for lattice Boltzmann equation},} \emph{Computers
  {\&} Mathematics with Applications}, Vol.~58, No.~5, 2009, pp. 903--913.
\newblock \doi{10.1016/j.camwa.2009.02.013}.

\bibitem[{Najafi-Yazdi and Mongeau(2012)}]{najafi-comflu-2012}
Najafi-Yazdi, A., and Mongeau, L., \enquote{{An absorbing boundary condition
  for the lattice Boltzmann method based on the perfectly matched layer},}
  \emph{Computers {\&} Fluids}, Vol.~68, 2012, pp. 203--218.
\newblock \doi{10.1016/j.compfluid.2012.07.017}.

\bibitem[{Shao and Li(2018)}]{shao-jtca-2018}
Shao, W., and Li, J., \enquote{An Absorbing Boundary Condition Based on
  Perfectly Matched Layer Technique Combined with Discontinuous Galerkin
  Boltzmann Method for Low Mach Number Flow Noise,} \emph{Journal of
  Theoretical and Computational Acoustics}, Vol.~26, No.~04, 2018, p. 1850011.
\newblock \doi{10.1142/s2591728518500111}.

\bibitem[{Poinsot and Lele(1992)}]{poinsot-jocp-1992}
Poinsot, T., and Lele, S., \enquote{{Boundary conditions for direct simulations
  of compressible viscous flows},} \emph{Journal of Computational Physics},
  Vol. 101, No.~1, 1992, pp. 104--129.
\newblock \doi{10.1016/0021-9991(92)90046-2}.

\bibitem[{Colonius et~al.(1993)Colonius, Lele, and Moin}]{colonius-aiaa-1993}
Colonius, T., Lele, S.~K., and Moin, P., \enquote{Boundary conditions for
  direct computation of aerodynamic sound generation,} \emph{AIAA Journal},
  Vol.~31, No.~9, 1993, pp. 1574--1582.
\newblock \doi{10.2514/3.11817}.

\bibitem[{Freund(1997)}]{freund-aiaa-1997}
Freund, J.~B., \enquote{Proposed Inflow/Outflow Boundary Condition for Direct
  Computation of Aerodynamic Sound,} \emph{AIAA Journal}, Vol.~35, No.~4, 1997,
  pp. 740--742.
\newblock \doi{10.2514/2.167}.

\bibitem[{Chen and Xuan(2020)}]{chen-pre-2020}
Chen, Y., and Xuan, Y., \enquote{Lattice Boltzmann approach for near-field
  thermal radiation,} \emph{Physical Review E}, Vol. 102, No.~4, 2020.
\newblock \doi{10.1103/physreve.102.043308}.

\bibitem[{Rajamuni et~al.(2023)Rajamuni, Liu, Wang, Ravi, Young, Lai, and
  Tian}]{rajamuni-preprint-2023}
Rajamuni, M.~M., Liu, Z., Wang, L., Ravi, S., Young, J., Lai, J. C.~S., and
  Tian, F.-B., \enquote{An Immersed Boundary-Regularised Lattice Boltzmann
  Method for Modelling Fluid-Structure-Acoustics Interactions Involving Large
  Deformation,} 2023.
\newblock \doi{10.2139/ssrn.4624793}.

\bibitem[{Chen et~al.(2023)Chen, Yang, and Shan}]{chen-jocp-2023}
Chen, X., Yang, K., and Shan, X., \enquote{{Characteristic boundary condition
  for multispeed lattice Boltzmann model in acoustic problems},} \emph{Journal
  of Computational Physics}, 2023, p. 112302.
\newblock \doi{10.1016/j.jcp.2023.112302}.

\bibitem[{Shan et~al.(2006)Shan, Yuan, and Chen}]{shan-jofm-2006}
Shan, X., Yuan, X.-F., and Chen, H., \enquote{{Kinetic theory representation of
  hydrodynamics: a way beyond the Navier-Stokes equation},} \emph{Journal of
  Fluid Mechanics}, Vol. 550, 2006, pp. 413--441.
\newblock \doi{10.1017/S0022112005008153}.

\bibitem[{Shan(2010)}]{shan-pre-2010}
Shan, X., \enquote{{General solution of lattices for Cartesian lattice
  Bhatanagar-Gross-Krook models},} \emph{Physical Review E}, Vol.~81, 2010, p.
  036702.
\newblock \doi{10.1103/PhysRevE.81.036702}.

\bibitem[{Kr\"uger et~al.(2017)Kr\"uger, Kusumaatmaja, Kuzmin, Shardt, Silva,
  and Viggen}]{kruger-book-2017}
Kr\"uger, T., Kusumaatmaja, H., Kuzmin, A., Shardt, O., Silva, G., and Viggen,
  E.~M., \emph{{The Lattice Boltzmann Method}}, Springer International
  Publishing, 2017.
\newblock \doi{10.1007/978-3-319-44649-3}.

\bibitem[{Succi(2018)}]{succi-book-2018}
Succi, S., \emph{{The Lattice Boltzmann Equation: For Complex States of Flowing
  Matter}}, OUP Oxford, 2018.
\newblock \doi{10.1093/oso/9780199592357.001.0001}.

\bibitem[{Bhatnagar et~al.(1954)Bhatnagar, Gross, and
  Krook}]{bhatnagar-pr-1954}
Bhatnagar, P.~L., Gross, E.~P., and Krook, M., \enquote{{A Model for Collision
  Processes in Gases. Amplitude Processes in Charged and Neutral One-Component
  Systems},} \emph{Physical review}, Vol.~94, No.~3, 1954, pp. 511--525.
\newblock \doi{10.1103/PhysRev.94.511}.

\bibitem[{Chapman and Cowling(1990)}]{chapman-book-1990}
Chapman, S., and Cowling, T.~G., \emph{{The Mathematical Theory of Non-Uniform
  Gases, 3rd ed}}, 3\textsuperscript{rd} ed., Cambridge Mathematical Library,
  Cambridge University Press, 1990.
\newblock \urlprefix\url{https://books.google.de/books?id=y2Yyy798WzIC}.

\bibitem[{Shan(2016)}]{shan-jocs-2016}
Shan, X., \enquote{{The mathematical structure of the lattices of the lattice
  Boltzmann method},} \emph{Journal of Computational Science}, Vol.~17, 2016,
  pp. 475--481.
\newblock \doi{10.1016/j.jocs.2016.03.002}.

\bibitem[{Philippi et~al.(2006)Philippi, Hegele, dos Santos, and
  Surmas}]{philippi-pre-2006}
Philippi, P.~C., Hegele, L.~A., dos Santos, L. O.~E., and Surmas, R.,
  \enquote{{From the continuous to the lattice Boltzmann equation: The
  discretization problem and thermal models},} \emph{Physical Review E},
  Vol.~73, No.~5, 2006.
\newblock \doi{10.1103/physreve.73.056702}.

\bibitem[{Modave et~al.(2014)Modave, Delhez, and Geuzaine}]{modave-nme-2014}
Modave, A., Delhez, E., and Geuzaine, C., \enquote{{Optimizing perfectly
  matched layers in discrete contexts},} \emph{International Journal for
  Numerical Methods in Engineering}, Vol.~99, No.~6, 2014, pp. 410--437.
\newblock \doi{10.1002/nme.4690}.

\bibitem[{Klass et~al.(2023)Klass, Gabbana, and Bartel}]{klass-cicp-2023}
Klass, F., Gabbana, A., and Bartel, A., \enquote{{A Characteristic Boundary
  Condition for Multispeed Lattice Boltzmann Methods},} \emph{Communications in
  Computational Physics}, Vol.~33, No.~1, 2023, pp. 101--117.
\newblock \doi{10.4208/cicp.oa-2022-0052}.

\bibitem[{Klass et~al.(2024)Klass, Gabbana, and Bartel}]{klass-camwa-2024}
Klass, F., Gabbana, A., and Bartel, A., \enquote{{Characteristic boundary
  condition for thermal lattice Boltzmann methods},} \emph{Computers \&
  Mathematics with Applications}, Vol. 157, 2024, pp. 195--208.
\newblock \doi{10.1016/j.camwa.2023.12.033}.

\bibitem[{Thompson(1987)}]{thompson-jocp-1987}
Thompson, K.~W., \enquote{{Time dependent boundary conditions for hyperbolic
  systems},} \emph{Journal of Computational Physics}, Vol.~68, No.~1, 1987, pp.
  1--24.
\newblock \doi{10.1016/0021-9991(87)90041-6}.

\bibitem[{Yoo et~al.(2005)Yoo, Wang, Trouv{\'{e}}, and Im}]{yoo-ctan-2005}
Yoo, C.~S., Wang, Y., Trouv{\'{e}}, A., and Im, H.~G., \enquote{{Characteristic
  boundary conditions for direct simulations of turbulent counterflow flames},}
  \emph{Combustion Theory and Modelling}, Vol.~9, No.~4, 2005, pp. 617--646.
\newblock \doi{10.1080/13647830500307378}.

\bibitem[{Jung et~al.(2015)Jung, Seo, and Yoo}]{jung-jocp-2015}
Jung, N., Seo, H.~W., and Yoo, C.~S., \enquote{{Two-dimensional characteristic
  boundary conditions for open boundaries in the lattice Boltzmann methods},}
  \emph{Journal of Computational Physics}, Vol. 302, 2015, pp. 191--199.
\newblock \doi{10.1016/j.jcp.2015.08.044}.

\bibitem[{Heubes et~al.(2014)Heubes, Bartel, and Ehrhardt}]{heubes-jcam-2014}
Heubes, D., Bartel, A., and Ehrhardt, M., \enquote{{Characteristic boundary
  conditions in the lattice Boltzmann method for fluid and gas dynamics},}
  \emph{Journal of Computational and Applied Mathematics}, Vol. 262, 2014, pp.
  51--61.
\newblock \doi{10.1016/j.cam.2013.09.019}.

\bibitem[{Klass et~al.(2021)Klass, Gabbana, and Bartel}]{klass-jocs-2021}
Klass, F., Gabbana, A., and Bartel, A., \enquote{{A non-equilibrium bounce-back
  boundary condition for thermal multispeed LBM},} \emph{Journal of
  Computational Science}, Vol.~53, 2021, p. 101364.
\newblock \doi{10.1016/j.jocs.2021.101364}.

\bibitem[{Yoo and Im(2007)}]{yoo-ctam-2007}
Yoo, C.~S., and Im, H.~G., \enquote{{Characteristic boundary conditions for
  simulations of compressible reacting flows with multi-dimensional, viscous
  and reaction effects},} \emph{Combustion Theory and Modelling}, Vol.~11,
  No.~2, 2007, pp. 259--286.
\newblock \doi{10.1080/13647830600898995}.

\bibitem[{Bedrunka et~al.(2021)Bedrunka, Wilde, Kliemank, Reith, Foysi, and
  Kr{\"a}mer}]{bedrunka-hpc-2021}
Bedrunka, M.~C., Wilde, D., Kliemank, M., Reith, D., Foysi, H., and Kr{\"a}mer,
  A., \enquote{Lettuce: PyTorch-Based Lattice Boltzmann Framework,} Springer
  International Publishing, Cham, 2021, pp. 40--55.
\newblock \doi{10.1007/978-3-030-90539-2_3}.

\bibitem[{Corbetta et~al.(2023)Corbetta, Gabbana, Gyrya, Livescu, Prins, and
  Toschi}]{corbetta-epje-2023}
Corbetta, A., Gabbana, A., Gyrya, V., Livescu, D., Prins, J., and Toschi, F.,
  \enquote{Toward learning Lattice Boltzmann collision operators,} \emph{The
  European Physical Journal E}, Vol.~46, No.~3, 2023, p.~10.
\newblock \doi{10.1140/epje/s10189-023-00267-w}.

\end{thebibliography}

\end{document}